\documentclass[prd,nofootinbib,preprint,groupedaddress,secnumarabic]{revtex4}
\pdfoutput=1
\usepackage{amsmath}
\usepackage[pdftex]{graphicx}
\usepackage{color}

\ifx\pdfoutput\undefined
\usepackage[dvips,bookmarks=false]{hyperref}	
\else
\usepackage{hyperref}	
\fi
\hypersetup{colorlinks,bookmarksopen,bookmarksnumbered,citecolor=black,
linkcolor=black,pdfstartview=FitH,urlcolor=black}

\newcommand{\be}{\begin{equation}}
\newcommand{\ee}{\end{equation}}
\newcommand{\bea}{\begin{eqnarray}}
\newcommand{\eea}{\end{eqnarray}}
\newcommand{\capdef}{}
\newcommand{\mycaption}[2][\capdef]{\renewcommand{\capdef}{#2}%
       \caption[#1]{{\footnotesize #2}}}
\makeatletter
\renewcommand{\fnum@table}{\textbf{\tablename~\thetable}}
\renewcommand{\fnum@figure}{\textbf{\figurename~\thefigure}}
\makeatother

\newcommand{\rem}[1]{}

\begin{document}

\title{Dark Matter and Neutrino Masses from\\  
Global $U(1)_{B-L}$ Symmetry Breaking\vspace*{1cm}}

\author{Manfred Lindner}
\email{lindner_at_mpi-hd.mpg.de}

\author{Daniel Schmidt}
\email{daniel.schmidt_at_mpi-hd.mpg.de}

\author{Thomas Schwetz}
\email{schwetz_at_mpi-hd.mpg.de}

\affiliation{\vspace*{0.5cm}Max-Planck-Institut f\"{u}r Kernphysik, Saupfercheckweg
1, 69117 Heidelberg, Germany\vspace*{1cm}}

\begin{abstract}
\vspace*{0.5cm} We present a scenario where neutrino masses and Dark Matter
are related due to a global $U(1)_{B-L}$ symmetry. Specifically we consider
neutrino mass generation via the Zee--Babu two-loop mechanism, augmented by
a scalar singlet whose VEV breaks the global $U(1)_{B-L}$ symmetry.  In
order to obtain a Dark Matter candidate we introduce two Standard Model
singlet fermions. They form a Dirac particle and are stable because
of a remnant $Z_2$ symmetry. Hence, in this model the stability of Dark
Matter follows from the global $U(1)_{B-L}$ symmetry. We discuss the Dark
Matter phenomenology of the model, and compare it to similar models based on
gauged $U(1)_{B-L}$. We argue that in contrast to the gauged versions, the
model based on the global symmetry does not suffer from severe constraints
from $Z'$ searches.
\end{abstract}

\maketitle

\section{Introduction}

Observations of neutrino oscillations~\cite{Fukuda:1998mi, Ahmad:2002jz,
Araki:2004mb, Adamson:2008zt} imply that the Standard Model (SM) has to be
extended in order to give mass to neutrinos. The scale for neutrino masses
set by the larger of the two measured mass-squared differences is $m_\nu
\sim 0.05$~eV. Another hint for physics beyond the SM comes from
various cosmological and astrophysical observations, which require the
existence of Dark Matter (DM), presumably a new kind of particle beyond the known
ones. From WMAP CMB measurements combined with other cosmological observables
one obtains for the fraction of the Dark Matter density ($\Omega_{\rm DM}$)
to the total energy density of the Universe $\Omega_{\rm DM}
h^2=0.1123\pm0.0035$ while for normal baryonic matter one has only about
20\% thereof: $\Omega_b h^2=0.0226\pm0.0005$~\cite{Komatsu:2010fb}. Here $h$
is the Hubble parameter in units of 100~km~s$^{-1}$~Mpc$^{-1}$, with $h^2
\approx 0.5$. A natural candidate for a Dark Matter particle is a weakly
interacting massive particle (WIMP), since a stable particle with mass at
the ``weak scale'' (of order 100~GeV) and a ``typical'' annihilation cross
section acquires a relic abundance through thermal freeze-out in the early
Universe with $\Omega_{\rm DM}$ in rough agreement with observations.

At first sight the two scales indicated by neutrino mass and WIMP Dark
Matter are vastly different, by about 12 orders of magnitude. Nevertheless,
it is tempting to consider a common origin of them. In particular, it might
be possible that neutrino masses are generated by physics at the TeV scale.
In such a case a connection between the neutrino mass mechanism and Dark
Matter may exist, and --- in the best of all worlds --- both of them might
be testable at LHC. In this Letter we present a common framework linking
these two mysteries and offering characteristic signatures at LHC. 

There exist a number of publications which relate neutrino mass and Dark
Matter; a few examples can be found in refs.~\cite{Krauss:2002px,
Cheung:2004xm, Ma:2006km, Babu:2007sm, Gu:2008yj, Sahu:2008aw,
Hambye:2009pw, Farzan:2009ji, Aoki:2008av, Aoki:2009vf, Ma:2009gu, Okada:2010wd, Li:2010rb,
Farzan:2010mr, Meloni:2010sk, Hirsch:2010ru, Adulpravitchai:2010wd,
Kanemura:2011vm, Chang:2011kv, Adulpravitchai:2011ei}.  We provide another
interesting scenario where a global $U(1)_{B-L}$ plays a central role.  For
the neutrino mass mechanism we depart from the Zee--Babu
model~\cite{Zee:1985rj,Zee:1985id,Babu:1988ki}, where two $SU(2)_{\rm L}$
singlet scalars are introduced, one single and one double charged, and
neutrino masses are generated at two-loop level. Phenomenological studies of
this model have been performed, e.g., in refs.~\cite{Babu:2002uu,
AristizabalSierra:2006gb, Nebot:2007bc, Ohlsson:2009vk}. We extend this
model in a simple way by adding one scalar and two fermionic singlets.  In
the spirit of Majoron models (see e.g., \cite{Chikashige:1980ui,
Gelmini:1980re}) we impose the conservation of lepton number, or actually
$B-L$ (baryon minus lepton) number at the quantum level. The corresponding
$U(1)$ symmetry is spontaneously broken by the non-zero vacuum expectation
value (VEV) of the scalar singlet \cite{Chang:1988aa}. This induces the
tri-linear term in the scalar potential necessary for neutrino mass
generation and provides a (Dirac) mass for the Dark Matter. This
leads to a common source for lepton number breaking (and hence the
generation of a Majorana neutrino mass) and Dark Matter. 

Any model for Dark Matter has to address the question of why the DM particle
is stable on cosmological time scales.  This is often achieved by
introducing a $Z_2$ symmetry which protects the DM particle, a famous
example being R-parity in Supersymmetric theories. The adhoc postulation of
such a $Z_2$ symmetry may appear artificial, but our model provides an
example where the $Z_2$ symmetry emerges naturally as an unbroken remnant of
some larger symmetry of the theory, namely the global $U(1)_{B-L}$ symmetry.
This can be considered as a specific example of a more general class of
models, where the dark sector respects a global $U(1)$ (accidental or
imposed), which gets spontaneously broken in such a way that a $Z_2$
symmetry remains unbroken and guarantees the stability of Dark Matter. 

The plan of the Letter is as follows. We start by discussing the model in
sec.~\ref{sec:model}.  In sec.~\ref{sec:B-L} we present the Zee--Babu model
extended with the global $U(1)_{B-L}$ symmetry and in sec.~\ref{sec:scalar}
we discuss the scalar sector of the model. This includes a discussion of
various consequences of the massless Goldstone boson, the Majoron, from the
breaking of the global $U(1)$ symmetry. We comment on implications for BBN,
Higgs searches at colliders, 
and baryon genesis.
In sec.~\ref{sec:DM} we introduce the DM
fermions to the model. Secs.~\ref{sec:relic} and \ref{sec:DD} deal with the
relic DM abundance and direct detection prospects, respectively. 
In secs.~\ref{sec:indir} and \ref{sec:self} we briefly comment on
indirect detection signals and on Majoron induced DM self-interactions, respectively. 
A general discussion follows in Sec.~\ref{sec:discussion}, where we also compare our model based on the global $U(1)_{B-L}$ to models using a gauged $U(1)_{B-L}$ symmetry.

\section{The model}
\label{sec:model}

\subsection{The Zee--Babu model and spontaneous breaking of $B-L$}
\label{sec:B-L} 

The scalar sector of the original Zee--Babu
model~\cite{Zee:1985rj,Zee:1985id,Babu:1988ki} contains in addition to the
Standard Model Higgs doublet two complex SU(2)$_L$ singlet scalars: a singly charged
scalar $h^+$ and a doubly charged scalar $k^{++}$. Then $h^+$
couples to left-handed lepton doublets $L$ and $k^{++}$ couples to
right-handed leptons $\ell_R$ such that the contribution to the Lagrangian is
\begin{equation}
\mathcal{L}_{\rm lept}=f_{ij} \, L_{i}^{T} \mathbf{C}^{-1}\, i\sigma_2 \,
L_j  h^+
\,+\,
g_{ij} \, \ell_{Ri}^T \mathbf{C}^{-1} \ell_{Rj} \,k^{++} + {\rm h.c.}
\label{eq:BabuL} \,,
\end{equation}
where $\mathbf{C}$ is the charge conjugation matrix, $i,j$ label flavour
indices, and the Yukawa couplings $f$ and $g$ are antisymmetric and
symmetric, respectively. If both, $h^+$ and $k^{++}$, are assigned lepton
number $-2$, lepton number is conserved by these interactions and therefore
the theory respects a global $U(1)$ symmetry associated with $B-L$
conservation.\footnote{Note that we only write down the leptonic part of the Lagrangian. 
In a complete theory baryon number would have to be considered as
well and sphaleron processes would break $B+L$, while $B-L$ is preserved.} ~
In order to break lepton number and generate a Majorana mass
term for neutrinos, usually a tri-linear term is introduced in the scalar
potential, $\mu k^{++}h^{-}h^{-}$, which breaks $B-L$ explicitly.  Here we
forbid such a term by demanding that $B-L$ is an unbroken symmetry of the
Lagrangian. Instead we introduce a complex scalar $\varphi$, which is
singlet under the SM gauge group and has lepton number $-2$. Hence the VEV
of $\varphi$ breaks $B-L$ spontaneously~\cite{Chang:1988aa}. The scalar potential contains a
term
\be\label{eq:Vmu}
V_\mu = \lambda_{\mu}\varphi k^{++}h^{-}h^{-} + {\rm h.c.} \,,
\ee
inducing the ``$\mu$-term'' once $\varphi$ acquires its VEV, with 
\begin{equation}\label{eq:VEV}
\mu = \lambda_\mu \frac{w}{\sqrt{2}} \,,\qquad
\langle \varphi \rangle = \frac{w}{\sqrt{2}} \,.
\end{equation}

Light neutrino masses are generated via a two-loop diagram, which yields
\begin{eqnarray}\label{eq:m}
(m_{\nu})_{ab} = 16 \mu f_{ac} m_c g^*_{cd} I_{cd} m_d f_{bd} \,,
\end{eqnarray}
where $m_c$ are charged lepton masses and $I_{cd}$ is a two-loop
integral \cite{McDonald:2003zj}. Data from the LEP and Tevatron
colliders require that the masses of charged scalars $m_h$ and $m_k$ are typically
larger than ${\cal O}(100~{\rm GeV})$ \cite{Rentala:2011mr}. Hence we can neglect the masses of
charged leptons compared to them. In this case, one finds
\begin{equation}\label{eq:I}
I_{cd} \approx I = \frac{1}{(16\pi^2)^2}\frac{1}{M^2}\frac{\pi^2}{3}
\tilde I \left(\frac{m_k^2}{m_h^2}\right) \,,
\end{equation}
where $M={\rm max}(m_k, m_h)$ and $\tilde I(r)$ is a dimensionless function
of order unity~\cite{Nebot:2007bc}. Using eq.~\eqref{eq:I}, the light
neutrino mass matrix becomes
\begin{eqnarray}\label{eq:mnu}
m_{\nu} \simeq \frac{1}{48\pi^2} \frac{\mu}{M^2} \, \tilde I \, f
D_\ell g^\dagger D_\ell f^T \ ,
\end{eqnarray}
where the matrix $D_\ell = {\rm diag}(m_e,m_\mu,m_\tau)$ contains the
charged-lepton masses. Due to the
antisymmetry
of $f$, we have
$\det m_\nu = 0$, and therefore, one of the neutrinos is massless 
as long as 
higher-order corrections are not considered. The neutrino phenomenology as
well as other signatures of the model have been studied, e.g., in
refs.~\cite{Babu:2002uu, AristizabalSierra:2006gb, Nebot:2007bc,
Ohlsson:2009vk}.  

Light neutrino masses are suppressed by the heavy scalar masses and
proportional to the scale of lepton-number violation $\mu$, set by the $B-L$
breaking scale $w$, see eq.~\eqref{eq:VEV}. Assuming $\mu \sim M \sim
\Lambda$, we see from eq.~\eqref{eq:mnu} that for $m_\nu\sim 0.1$~eV the
scale $\Lambda$ of new physics has to be of order 1~TeV if we demand that
$f\sim g\sim0.05$. This scale $\Lambda$ for generating neutrino masses is
much below the scale of seesaw models due to the two loop suppression factor
$1/(16\pi^2)^2 \sim 10^{-4}$. Hence, $\Lambda$ is in the range of the LHC
and may thus be probed soon. In particular, the double charged scalar
provides a clean signature at colliders, via the decay into two like-sign
leptons. Furthermore, the exchange of the charged scalars leads to enhanced
signals in charged lepton flavour violation ($\mu\to e\gamma$ or $\mu\to
3e$) with good prospects to be observed soon \cite{AristizabalSierra:2006gb,
Nebot:2007bc}.

\subsection{The scalar sector and the Majoron}
\label{sec:scalar}

The full scalar potential of the model is 
\begin{eqnarray}
V_{\rm scalar} = V_\mu &+&\mu_{1}^{2}\varphi^{\ast}\varphi+\mu_{2}^{2}\phi^{\dagger}\phi+
\mu_{3}^{2}k^{++}k^{--}+\mu_{4}^{2}h^{+}h^{-}\nonumber\\
&+&\lambda_{1}(\varphi^{\ast}\varphi)^{2}+\lambda_{2}(\phi^{\dagger}\phi)^{2}+
\lambda_{3}(k^{++}k^{--})^{2}+\lambda_{4}(h^{+}h^{-})^{2}\nonumber\\
&+&\lambda_{5}(\varphi^{\ast}\varphi)(\phi^{\dagger}\phi)+\lambda_{6}
(\varphi^{\ast}\varphi)(k^{++}k^{--})+\lambda_{7}(\varphi^{\ast}\varphi)(h^{+}h^{-})\nonumber\\
&+&\lambda_{8}(\phi^{\dagger}\phi)(k^{++}k^{--})+\lambda_{9}(\phi^{\dagger}\phi)(h^{+}h^{-})+
\lambda_{10}(k^{++}k^{--})(h^{+}h^{-})
\label{eq:pot}\,,
\end{eqnarray}
where $V_\mu$ is given in eq.~\eqref{eq:Vmu}, $\phi$ denotes the Standard
Model Higgs doublet, $\mu_i$ are parameters of mass dimension one, and
$\lambda_i$ are dimensionless couplings. The $\lambda_5$ term in the
potential induces mixing between $\varphi$ and $\phi$. There are two massive
neutral scalars in the theory, with propagating mass states denoted by $H_1$
and $H_2$. They are related to the real part $\eta$ of $\varphi$ and to the
real part $\zeta$ of the neutral component of $\phi$ by 
\begin{equation}
  \begin{pmatrix}H_1\\H_2\end{pmatrix}=
  \begin{pmatrix}\cos\alpha&\sin\alpha\\-\sin\alpha&\cos\alpha\end{pmatrix}
  \begin{pmatrix}\eta\\\zeta\end{pmatrix}\,,
\end{equation}
where the mixing angle $\alpha$ is given by
\begin{equation}
  \tan2\alpha=\frac{\lambda_{5}wv}{\lambda_{1}w^{2}-\lambda_{2}v^{2}}\,,
\end{equation}
and $v$ denotes the VEV of the Higgs doublet $\phi$. The masses of
$H_1$ and $H_2$ are 
\be
m_{H_{1,2}}^2 = \lambda_{1}w^{2} + \lambda_{2}v^{2} \pm
\sqrt{\left(\lambda_{1}w^{2}-\lambda_{2}v^{2}\right)^{2}+\lambda_{5}^2w^{2}v^{2}} \,.
\ee
The parameters $\mu_{1,2}$ in eq.~\eqref{eq:pot} can be eliminated
with the help of the minimum condition for the potential.
Therefore, the neutral scalar phenomenology depends only on three independent
parameters (in addition to the VEVs $v,w$), which can be chosen to be either
$(\lambda_1,\lambda_2,\lambda_5)$ or alternatively, $(m_{H_1},
m_{H_2}, \alpha)$. 

Via Goldstone's theorem \cite{Goldstone:1961eq} a massless scalar has to
appear in the spectrum due to the spontaneous breaking of the global $U(1)$
symmetry. In our model the imaginary part of $\varphi$ remains massless. We
write
\be
\varphi = \frac{1}{\sqrt{2}}(w + \eta + i\rho) \,,
\ee
where $\eta = \sqrt{2} {\rm Re}(\varphi)$ and $\rho= \sqrt{2} {\rm
Im}(\varphi)$ are real scalar fields and $\rho$ is the Goldstone boson.
Since the VEV of $\varphi$ is responsible for lepton number breaking and the
generation of a Majorana mass for the Dark Matter (see later) we follow the
literature and call $\rho$ a Majoron. Note that $\varphi$ is a singlet under
the SM gauge group and therefore there is no direct coupling of $\rho$ to
the $Z$ boson like in triplet Majoron models \cite{Gelmini:1980re} which are
ruled out by the LEP measurement of the invisible $Z$ decay width. The
couplings of the Majoron $\rho$ to the Higgs mass eigenstates $H_1$ and
$H_2$ are obtained from the $\lambda_1$ and $\lambda_5$ terms of the
potential~\eqref{eq:pot}:
\begin{equation}\label{couplingMajoron}
\mathcal{L}_\rho = \frac{1}{2w}
  \left(m^2_{H_1}\cos\alpha\,H_1-m^2_{H_2}\sin\alpha\,H_2\right)\rho^2\,.
\end{equation} 

The massless Majoron contributes to the relativistic energy density in the
Universe, conventionally parameterized by the effective number of neutrino
species. One thermalized scalar contributes $\Delta n_\nu = 4/7$ neutrino
species, and therefore a thermal scalar degree of freedom is consistent with
bounds from Big Bang Nucleosynthesis (BBN), which are in the range of $\Delta n_\nu <
1.63$~\cite{Cyburt:2004yc} or $\Delta n_\nu < 1.2$~\cite{Mangano:2011ar} at
95\%~CL, depending on assumptions about the primordial abundances. Furthermore, in our model the Majoron typically decouples from the
plasma at temperatures above the QCD phase transition, where the effective
number of relativistic degrees of freedom is $\gtrsim 60$. Therefore, due to
the entropy production at the QCD phase transition the Majoron abundance
gets diluted and during  
BBN $\rho$ contributes only with
\begin{equation}
\Delta n_\nu \lesssim \frac{4}{7} \left(\frac{10.75}{60}\right)^{{4}/{3}}\approx0.06
\end{equation}
to the relativistic energy density, in good
agreement with the above
mentioned bounds.

The prospects for discovering  the charged scalars of the Zee--Babu model at the LHC
have been discussed in detail in \cite{Nebot:2007bc}. In
particular the double charged scalar has a production cross section at LHC
in the range of 50 to 0.1~fb for masses between 200~GeV and 1~TeV. Its decay
into two pairs of like-sign leptons provides an essentially background free
signature, see also \cite{Rentala:2011mr} for a recent analysis. Possible implications of the lepton flavour violating reactions mediated by the double charged scalar in supernova physics have been discussed in~\cite{Lychkovskiy:2010ue}.

In addition to the charged scalar signatures, the presence of the Majoron
will modify Higgs physics. As pointed out in \cite{Shrock:1982kd,
Joshipura:1992ua, Joshipura:1992hp} an important feature of Majoron models
are invisible Higgs decays $H \to \rho\rho$, see also \cite{Ghosh:2011qc} for a recent study. In the SM, the Higgs boson
dominantly decays into bottom pairs.  In our model the corresponding
branching ratios for the two Higgs mass eigenstates $H_1$ and $H_2$ are
weighted with the Higgs mixing angle $\alpha$:
\begin{subequations}
\begin{eqnarray}
\Gamma\left(H_1\rightarrow b\overline{b}\right)=\frac{3\sqrt{2}G_Fm^2_bm_{H_1}}{8\pi}\left(1-
\frac{4m^2_b}{m^2_{H_1}}\right)^{\frac{3}{2}}\,\sin^2\alpha\,,&&\\
\Gamma\left(H_2\rightarrow b\overline{b}\right)=\frac{3\sqrt{2}G_Fm^2_bm_{H_2}}{8\pi}\left(1-
\frac{4m^2_b}{m^2_{H_2}}\right)^{\frac{3}{2}}\,\cos^2\alpha\,.&&
\end{eqnarray}
\end{subequations}
The invisible decay modes into the Majoron $\rho$ are obtained
from the Lagrangian eq.~\eqref{couplingMajoron},
\begin{subequations}
\begin{eqnarray}
\Gamma\left(H_1\rightarrow\rho\rho\right)=\frac{\sqrt{2}G_F}{32\pi}m^3_{H_1}\left(\frac{v}{w}
\right)^2\cos^2\alpha\,,&&\\
\Gamma\left(H_2\rightarrow\rho\rho\right)=\frac{\sqrt{2}G_F}{32\pi}m^3_{H_2}\left(\frac{v}{w}
\right)^2\sin^2\alpha\,,&&
\end{eqnarray}
\end{subequations}
and therefore \cite{Joshipura:1992hp},
\begin{subequations}
\begin{eqnarray}
  \frac{\Gamma\left(H_1\rightarrow\rho\rho\right)}
       {\Gamma\left(H_1\rightarrow b\overline{b}\right)} = 
  \frac{1}{12}\left(\frac{m_{H_1}}{m_b}\right)^2\left(\frac{v}{w}\right)^2\cot^2\alpha
       \left(1-\frac{4m^2_b}{m^2_{H_1}}\right)^{-\frac{3}{2}}\label{part1}
  \approx 250 \left(\frac{m_{H_1}}{w}\right)^2 \cot^2\alpha \,,&&\\
  \frac{\Gamma\left(H_2\rightarrow\rho\rho\right)}
       {\Gamma\left(H_2\rightarrow b\overline{b}\right)} = 
  \frac{1}{12}\left(\frac{m_{H_2}}{m_b}\right)^2\left(\frac{v}{w}\right)^2\tan^2\alpha
      \left(1-\frac{4m^2_b}{m^2_{H_2}}\right)^{-\frac{3}{2}}\label{part2}
  \approx 250 \left(\frac{m_{H_2}}{w}\right)^2 \tan^2\alpha \,.&&
\end{eqnarray}
\end{subequations}
As we will see in the following, typically $m_{H_{1,2}} \sim w$, and
therefore, depending on the Higgs mixing $\alpha$, we expect that one or both
Higgses decay dominantly into invisible Majorons. In the literature, various
collider signatures have been studied for the identification of an invisible
decaying Higgs boson, for example associated production of $ZH$
\cite{Meisel:904271} and $t \overline{t}H$ \cite{refId}, or production in
weak vector boson fusion \cite{DiGirolamo:685420}.

\subsection{Baryon asymmetry}
\label{sec:genesis}

In our model we do not attempt to explain the generation of the baryon number of the universe, and rely on some unspecified mechanism beyond our model. However, in models with low-scale lepton number violation often the problem occurs that any pre-existing baryon asymmetry is washed out by sphaleron processes if the lepton number violating reactions are in equilibrium before sphaleron freeze out. In our model lepton number is a global symmetry, which gets broken spontaneously by the VEV $w$ of $\varphi$, somewhere between 200~GeV and 2~TeV, depending on parameter values, see fig.~\ref{fig:relic}. Hence, at temperatures above $w$ lepton number is conserved and a pre-existing asymmetry is not affected. Only at temperatures $T \lesssim w$ lepton number violating processes occur via reactions involving the term $\lambda_\mu w k^{++}h^-h^-$. If sphaleron processes are still in equilibrium at that time, a pre-existing baryon asymmetry will be washed out. Sphaleron processes freeze out at the electro-weak phase transition at temperatures $T\sim v$. Since generically $w\sim v$ both phase transitions (the $B-L$ as well as the electro-weak ones) happen at similar temperatures. It remains a quantitative question of how much of the baryon asymmetry is erased by the detailed interplay of both phase transitions, which is beyond the scope of our Letter. Note that there is some parameter space where $w$ is slightly less than $v$ (compare fig.~\ref{fig:relic}), and in this case $B-L$ would be broken only after sphaleron freeze out.


\section{Dark Matter}
\label{sec:DM}

Having a global $B-L$ symmetry which is motivated by neutrino 
masses, it is tempting to ask if it could play also a role in 
stabilizing Dark Matter. 
Therefore we introduce fermions $N_i$, which are singlets
under the SM gauge group, but charged under $U(1)_{B-L}$ in such a way
that the Yukawa interaction with the SM Higgs doublet $\phi$, $\bar
L_j \tilde \phi N_i$ $(\tilde\phi \equiv i\sigma_2\phi^*)$, is forbidden. Hence, our $N_i$
cannot have lepton number +1 and are therefore not ``right-handed
neutrinos'' in the conventional sense. Still we want the mass term for
$N_i$ to be generated by spontaneous lepton number breaking from the
term
\be\label{eq:Nmass}
\mathcal{L}_N = \frac{1}{2} h_{ij} \, \varphi \, N_i^T \mathbf{C}^{-1} N_j + {\rm h.c.} \,.
\ee

This can be achieved by introducing two $N$ fields, $N_1$ and
$N_2$, and assigning them lepton numbers $q_1$ and $q_2$, respectively, such that 
$q_1 + q_2 = 2$ with $q_1 \neq q_2 \neq 1$; for example
$q_1 = 1/2$ and $q_2 = 3/2$. Then no Yukawa
term with the lepton doublets is allowed and eq.~\eqref{eq:Nmass} leads to a
mass matrix
\be\label{eq:mchi} 
M_N = \left(\begin{array}{cc} 0 & m_\chi \\ m_\chi & 0 \end{array}\right)
\quad\text{with}\quad m_\chi = \lambda_\chi \, \frac{w}{\sqrt{2}} \,, 
\ee 
and $\lambda_\chi \equiv h_{12}$. Hence, $N_1$ and $N_2$ form a Dirac
particle, with a pair of degenerate mass eigenstates with mass $m_\chi$ and
opposite CP parity:
\be\label{eq:chiCoupl}
\chi_1 = \frac{1}{\sqrt{2}}(N_1 + N_2) \,,\qquad
\chi_2 = \frac{i}{\sqrt{2}}(N_1 - N_2) \,.
\ee
They are stable because of an accidental $Z_2$ symmetry which
emerges as an unbroken remnant of the global
$U(1)_{B-L}$.
\footnote{Note that there are more unbroken accidental symmetries in the Lagrangian. For example there is a $Z_3$ symmetry $N_1\to\omega N_1, N_2\to \omega^2N_2$ with $\omega^3=1$. Another example is an additional $U(1)$ symmetry with opposite charges for $N_1$ and $N_2$ but all other fields uncharged. Those accidental symmetries emerge due to the $B-L$ charge assignments of $N_{1,2}$ and they ensure that no Majorana mass term is generated for them after $B-L$ breaking.} 
The interaction eq.~\eqref{eq:Nmass} becomes diagonal in the $\chi_i$
fields:
\be\label{eq:Lchi}
\mathcal{L}_N = \frac{1}{2} \lambda_\chi \, \varphi 
 (\chi_1^T \mathbf{C}^{-1} \chi_1 + 
  \chi_2^T \mathbf{C}^{-1} \chi_2) + {\rm h.c.} \,.
\ee

Note that in general global symmetries and non-gauge discrete symmetries
(i.e., discrete symmetries which are not remnants of broken gauge
symmetries) are expected to be violated by quantum gravity effects,
e.g.~\cite{Giddings:1988cx}, see \cite{Banks:2010zn} for a recent
discussion.  Planck scale suppressed higher dimensional operators are
therefore expected in the low energy effective theory which would provide a
mass to the Majoron~\cite{Akhmedov:1992hi} and might also induce DM decay in
our model. If there is a dimension 6 operator suppressed by the Planck mass
$m_{pl}$ inducing DM decay, the corresponding lifetime exceeds the age of
the Universe. However, a Dark Matter decay operator of dimension 5 would be
problematic and should therefore be strongly suppressed or absent, since an
estimate for the lifetime yields
$\tau\sim{m_{pl}^2}/{m_{\chi}^3}\approx10^7$~s for $m_{\chi}\sim 100$~GeV
and there would be no DM left today. In the following we assume that Planck
scale suppressed operators induce DM decays only with lifetimes larger than
the age of the Universe.  We comment on the effects of a finite mass for the
Majoron in sec.~\ref{sec:discussion}

The only particle to which $\chi_i$ can couple is the scalar $\varphi$, with 
the coupling $\lambda_\chi$ which is related to the DM mass via the VEV $w$,
see eq.~\eqref{eq:mchi}. Furthermore, the coupling of $\varphi$ to the SM is
provided via the Higgs portal proportional to $\lambda_5$ (or the mixing
angle $\alpha$). Therefore, the two parameters $\lambda_\chi$ and $\alpha$
will play an important role for DM phenomenology, as we are going to discuss
in the following. 

\subsection{Relic Dark Matter abundance}
\label{sec:relic}

The relic DM density $\Omega_{\rm DM} h^2$ is determined
by the thermal freeze-out of $\chi_i$ in the early Universe. It is roughly
inversely proportional to the 
thermal average of the total annihilation cross section times the relative
velocity: $\Omega_{\rm DM} h^2 \simeq 3 \times 10^{-27} {\rm cm^3 s^{-1}} / 
\langle\sigma v_{\rm rel}\rangle$. Note that the two DM particles $\chi_1$ and $\chi_2$ have identical couplings to $\varphi$. Hence the annihilation cross sections are the same and they will contribute in equal parts to the total DM density. 
Two DM particles $\chi_i$ can annihilate into a quark--antiquark pair, into SM
gauge bosons, into the massive scalars $H_{1,2}$, into the Zee--Babu scalars $k^{++},h^+$, as well as in the Majoron
$\rho$. As examples we show the cross sections for annihilation via
$s$-channel $H_{1,2}$ exchange into $b\,\overline{b}$, $W^+W^-$ and $k^{++}k^{--}$, as well as the $t$- and $u$-channel annihilation into $\rho\rho$:
\begin{subequations}
\begin{eqnarray}
\sigma_{b\,\overline{b}} v_{\rm rel} &\approx& \lambda_\chi^2 \,  \sin^22\alpha \,
  \frac{y_b^2 v_{\rm rel}^2}{1024 \pi m_\chi} \cdot
  \frac{(m_{H_{1}}^2-m_{H_{2}}^2)^2}{(s-m_{H_{1}}^2)^2(s-m_{H_{2}}^2)^2} \,
  \left(s - 4 m_{b}^2\right)^{3/2} \label{crossHbb}\,,\\
\sigma_{WW} v_{\rm rel} &\approx& \lambda_\chi^2 \,  \sin^22\alpha \, 
  \frac{g^4 v^2 v_{\rm rel}^2}{2048 \pi m_\chi} \cdot 
  \frac{(m_{H_{1}}^2-m_{H_{2}}^2)^2}{(s-m_{H_{1}}^2)^2(s-m_{H_{2}}^2)^2}
  \sqrt{\left(s-4m_W^2\right)} \nonumber\\
  &&
  \qquad\qquad \times
  \left[1+\frac{1}{2}\left(\frac{s}{2m_W^2}-1\right)^2\right]\label{crossHWW}\,,\\
\sigma_{kk} v_{\rm rel} &=& \lambda_6^2 \,
  \frac{ m_\chi v_{\rm rel}^2}{32\pi} \, \frac{\left[(s-m_{H_2}^2)\cos\alpha-(s-m_{H_1}^2)\sin\alpha\right]^2}
  {(s-m^2_{H_1})^2 \, (s-m^2_{H_2})^2}\, \sqrt{s-4m^2_{k^{++}}} \label{eq:crosskk}\,,\\
\sigma_{\rho\rho}v_{\rm rel} &=& \lambda^4_\chi\, \frac{v_{\rm rel}^2}{1536\pi m^2_\chi}
\label{crossHrr}\,.
\end{eqnarray}
\end{subequations} 
We expanded $s = 4m^2_\chi / (1 - v_{\rm rel}^2 / 4)$ in $v_{\rm rel} \ll 1$
to show explicitly the velocity suppression of the annihilation, however we
keep $s$ in the kinematical terms as well as in the denominators showing the
resonant behavior of the $s$-channel cross section as a function of the DM
mass $m_\chi$. For simplicity we neglect here also the width of the
resonances, which are however included in the numerical calculations
presented below. In eq.~\eqref{crossHbb}, $y_b$ and $m_b$ are the $b$-quark
Yukawa coupling and mass, respectively.  The annihilation cross section of
DM into a $ZZ$ pair can be obtained from eq.~\eqref{crossHWW} by replacing
$g \to g' = g / (\sqrt{2} \cos\theta_W)$ and $m_W\rightarrow m_Z$. The
annihilation cross section into $h^+h^-$ can be obtained from
eq.~\eqref{eq:crosskk} by replacing $\lambda_6\to\lambda_7$ and $m_{k^{++}}
\to m_{h^+}$. Note that the full annihilation cross section
$\chi\chi\to\rho\rho$ receives also contributions from an $s$-channel
diagram, whereas eq.~\eqref{crossHrr} shows only the $t$- and $u$-channel
contribution for simplicity. This is the only case without the suppression
by the Higgs masses $m_{H_{1,2}}$. 


All annihilation cross sections are proportional to the coupling
$\lambda_\chi$, however, in eq.~\eqref{eq:crosskk} $\lambda_\chi$ appears
together with the VEV $w$ and has been absorbed into the DM mass via
eq.~\eqref{eq:mchi}. Annihilations into SM particles are controlled by the
Higgs portal, and therefore proportional to the mixing parameter
$\sin^22\alpha$. In contrast, annihilations into the Majoron and the
Zee--Babu scalars survive even in the case of no mixing ($\sin 2\alpha = 0$,
or $\lambda_5 = 0$).  Hence, these annihilation channels still provide a
mechanism to generate the correct relic abundance at the electro-weak scale
for $\sin 2 \alpha = 0$, despite of $\chi_i$'s being completely decoupled
from the SM.

\begin{figure}
\begin{center}
\begin{tabular}{ccc}
a)
\includegraphics[width=0.31\textwidth]{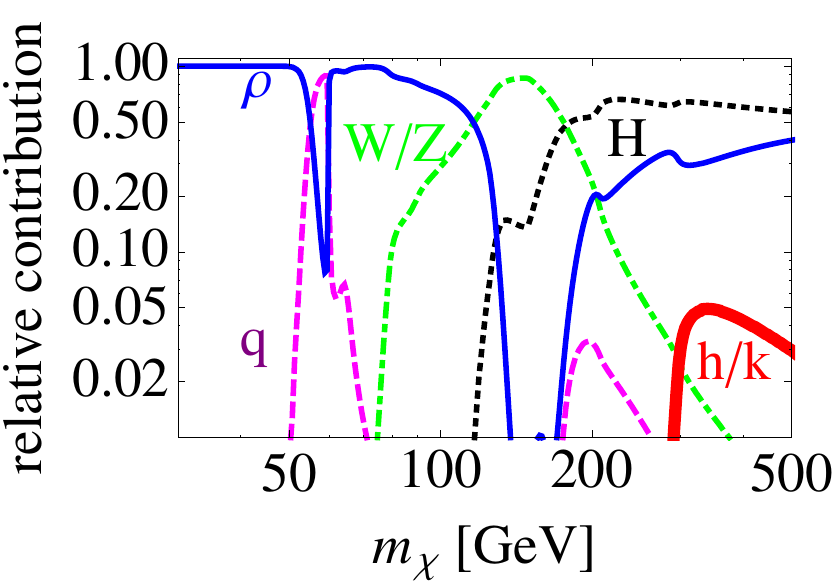} &
\includegraphics[width=0.31\textwidth]{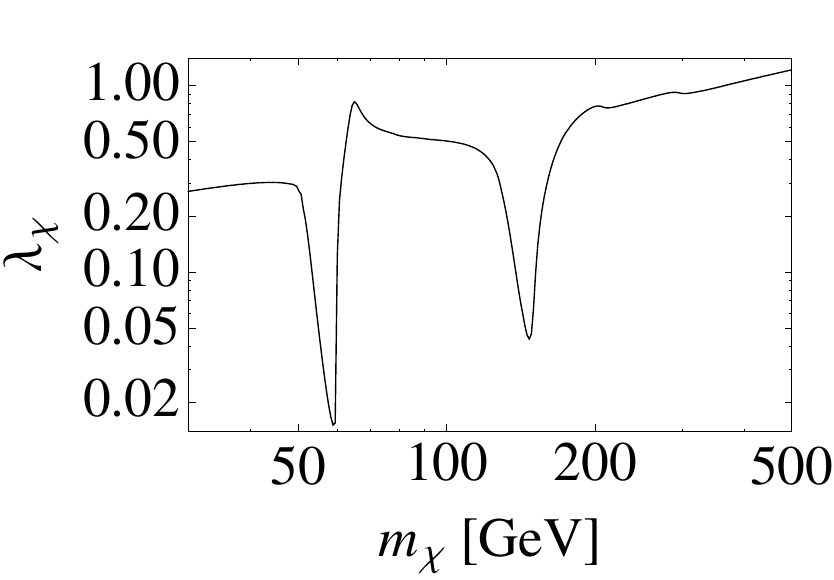} &
\includegraphics[width=0.31\textwidth]{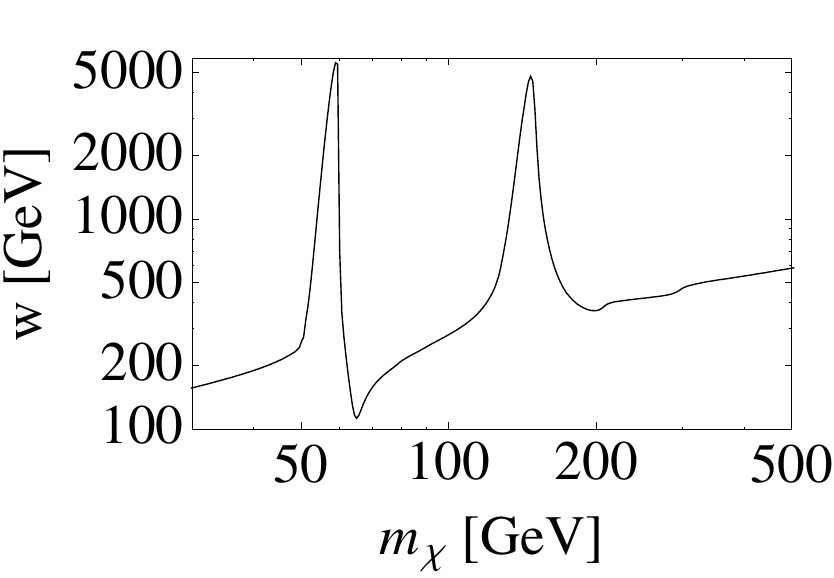} \\
b)
\includegraphics[width=0.31\textwidth]{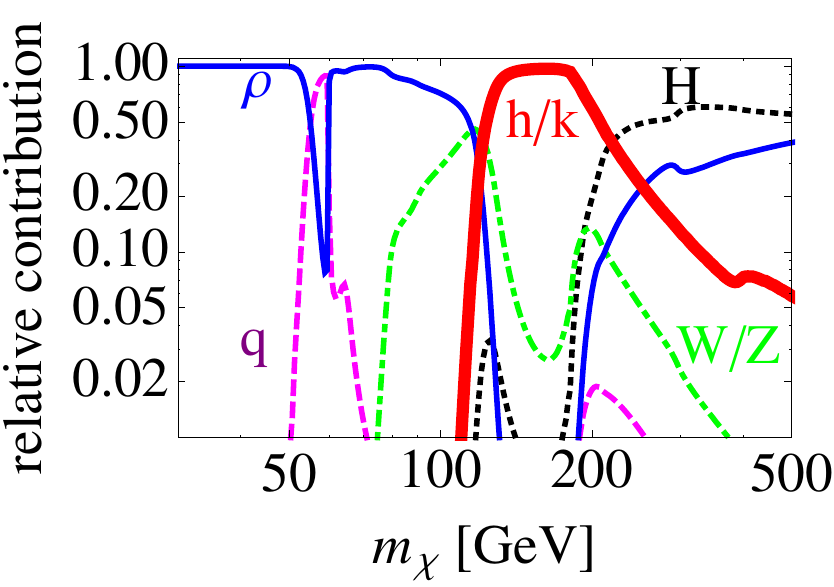} &
\includegraphics[width=0.31\textwidth]{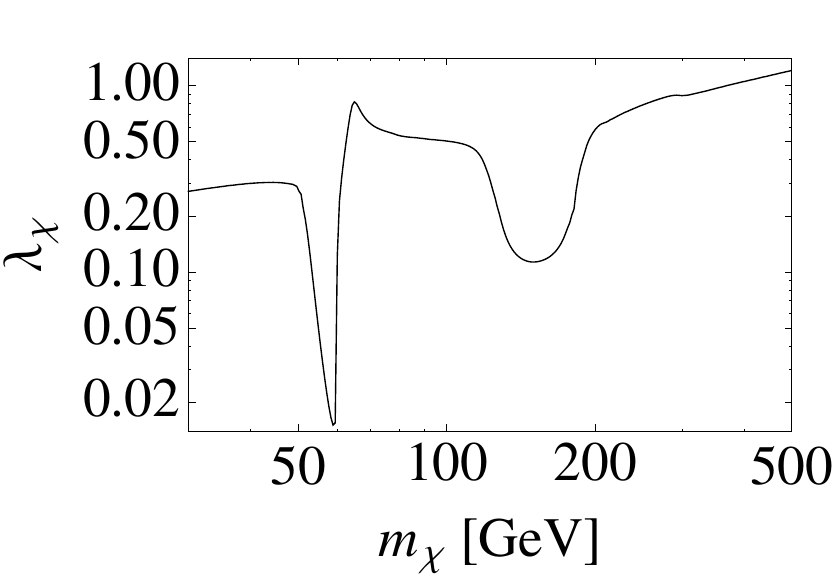} &
\includegraphics[width=0.31\textwidth]{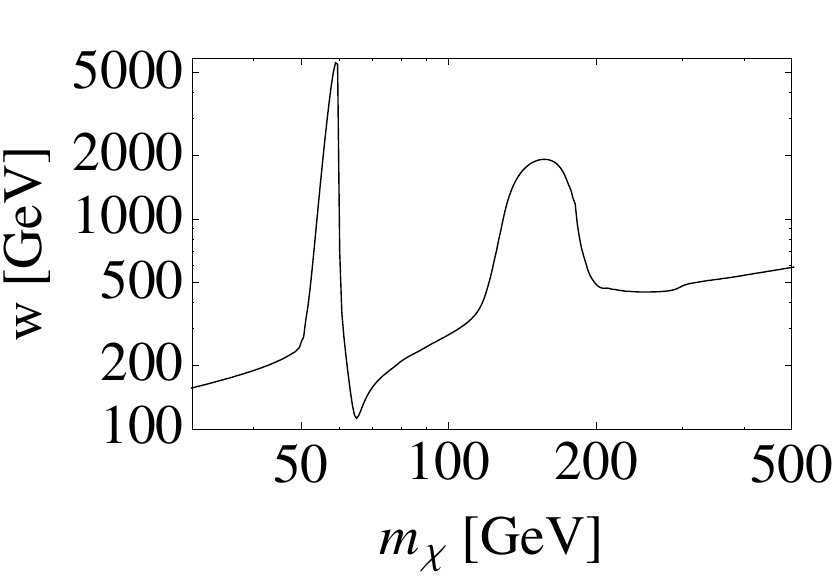}\\
c)
\includegraphics[width=0.31\textwidth]{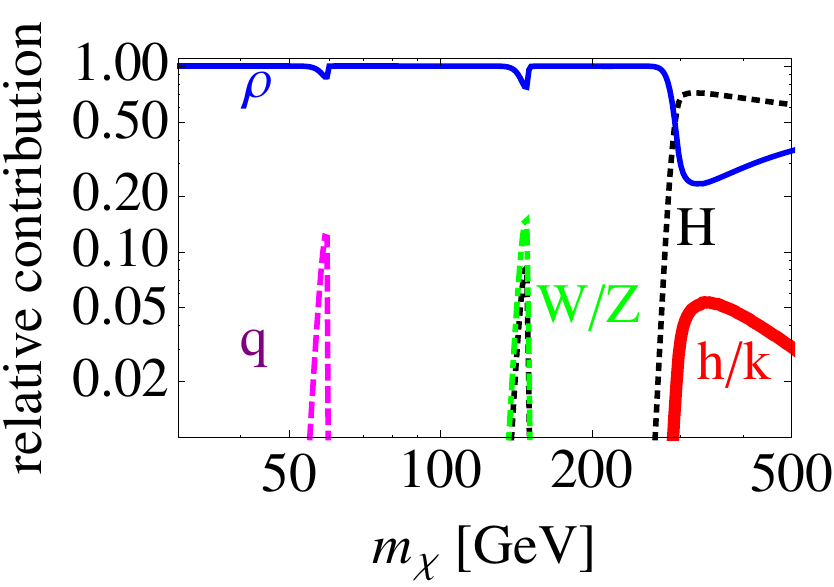} &
\includegraphics[width=0.31\textwidth]{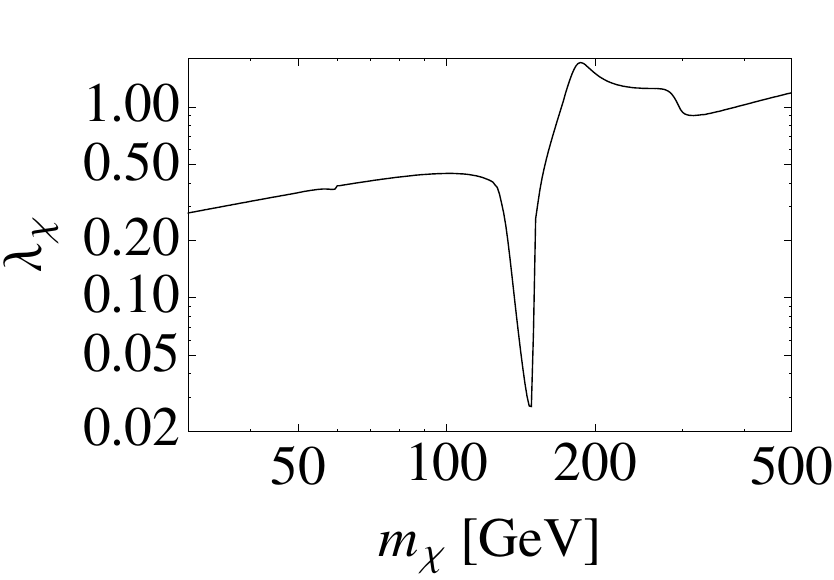} &
\includegraphics[width=0.31\textwidth]{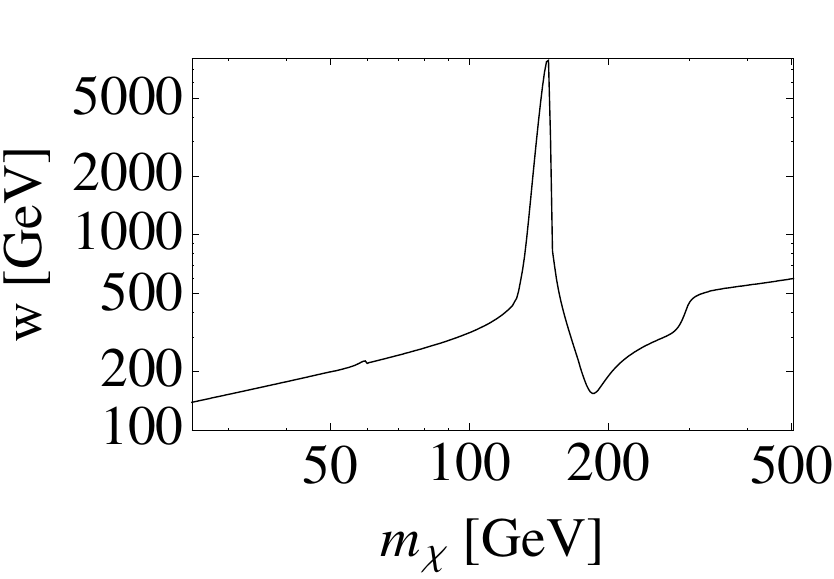} \\
d)
\includegraphics[width=0.31\textwidth]{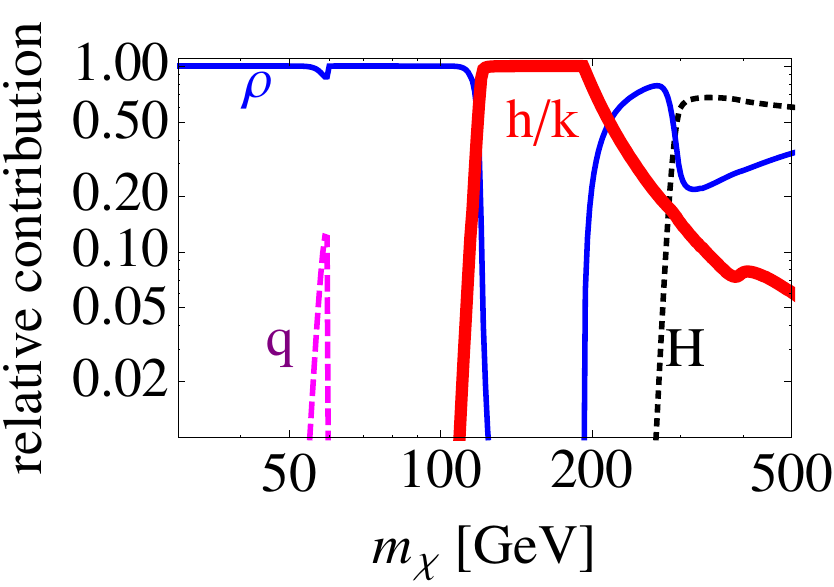} &
\includegraphics[width=0.31\textwidth]{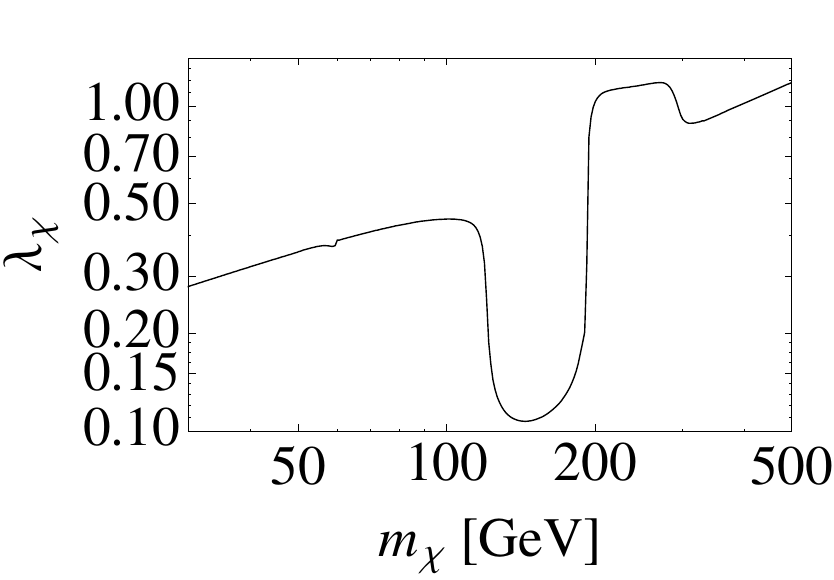} &
\includegraphics[width=0.31\textwidth]{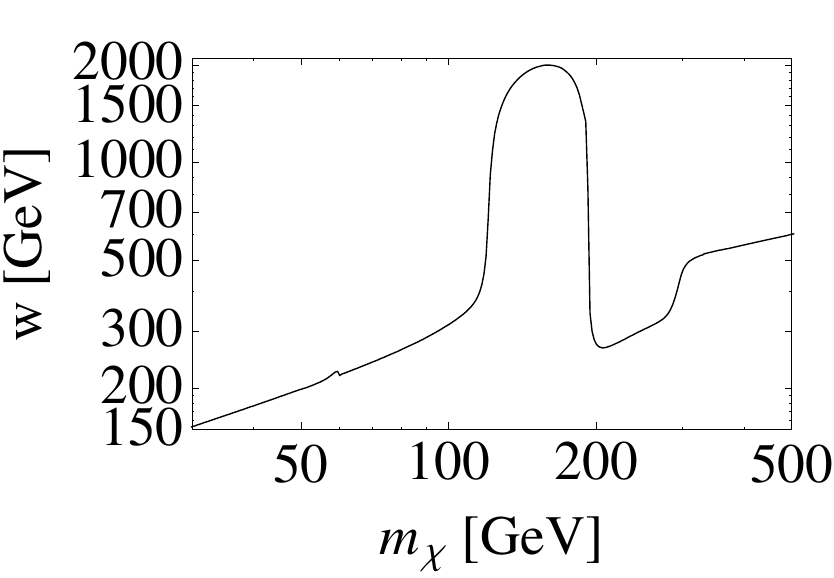}
\end{tabular}
  \mycaption{Left: Relative contribution of annihilation channels to the
  relic DM abundance. Shown are annihilations into $H_1$ and $H_2$ (``H''),
  quarks (``q''), W and Z bosons (``W/Z''), Majorons (``$\rho$'') and
  Zee--Babu scalars (``h/k''). Middle and right columns show the DM scalar
  coupling $\lambda_\chi$ and singlet VEV, respectively, which lead to the
  correct relic DM density. We assume $m_{H_1}=300$~GeV, $m_{H_2}=120$~GeV,
  and all Zee--Babu scalar couplings are set to unity. Rows a) and b) are
  for a Higgs mixing angle $\sin\alpha=0.7$, rows c) and d) for $\sin\alpha
  = 0.01$. Rows a) and c) are for Zee--Babu scalar masses $m_{h^+}=300$~GeV,
  $m_{k^{++}}=800$~GeV, rows b) and d) for $m_{h^+}=120$~GeV,
  $m_{k^{++}}=400$~GeV.} \label{fig:relic}
\end{center}
\end{figure}

We use the micrOMEGAs software package \cite{Belanger:2008sj} to calculate 
the relic density $\Omega_{\rm DM} = \Omega_{\chi_1} + \Omega_{\chi_2}$. For
a given set of parameters, the left column of plots in fig.~\ref{fig:relic}
shows the relative contribution of the various annihilation channels to
the relic abundance. The middle and the right panels of the figure show the
coupling $\lambda_\chi$ and the singlet VEV $w$ which are needed to obtain
the correct relic abundance. Note that for a given DM mass $m_\chi$,
$\lambda_\chi$ and $w$ are simply related by eq.~\eqref{eq:mchi}. We have
chosen representative values for the two scalar masses $m_{H_1} =
300$~GeV and $m_{H_2} = 120$~GeV, but we have verified that our conclusions
do not depend on this specific choice and hold within the full range of
``reasonable'' values for the Higgs masses. The couplings $\lambda_\mu,
\lambda_{3,4,6,7,8,9,10}$ have been set to one. Again we have checked that
random values in the range from $0.1$ to 1 give qualitative similar results.

The upper two rows of plots in fig.~\ref{fig:relic} correspond to a
relatively large Higgs mixing angle of $\sin\alpha = 0.7$ and there is a
large coupling of DM to the SM. We find that depending on the DM mass
various annihilation channels are important and the two resonances at
$m_\chi \simeq m_{H_{1,2}} /2$ are clearly visible. In the lower two rows we
use the same parameters but a small Higgs mixing $\sin\alpha = 0.01$. In
this case the mass eigenstate $H_1$ practically coincides with the singlet
$\varphi$ and therefore only the resonance corresponding to
$H_1$ exchange occurs. Furthermore, the coupling to SM particles is suppressed and
the relic density is provided only by annihilations into the massless Majoron or, if
kinematically accessible, into $H_1$ and the Zee--Babu scalars $k^{\pm\pm},
h^\pm$. 

In fig.~\ref{fig:relic} we compare also two assumptions on the masses of the
Zee--Babu scalars. For masses larger than the Higgs masses annihilations
into $k^{\pm\pm}$ and $h^\pm$ are sub-dominant, but if one of the charged
scalars is lighter than one of the neutral scalars they can dominate DM
annihilations, as visible in rows b) and d). Note also the modified shape of the
curves close to the resonance if the dominant annihilation
channel is into charged scalars. While the shape of the resonance of the
cross section itself remains the same, the different shape for 
$\lambda_\chi$ which gives the correct relic abundance follows from a
different dependence of the annihilation cross section on $\lambda_\chi$ and
$m_\chi$ visible from eq.~\eqref{eq:crosskk}, which shows an $m_\chi$
proportionality, compared to the typical $\lambda_\chi^2 / m_\chi$
proportionality of other channels.

The middle and right columns of plots in fig.~\ref{fig:relic} show that away
from the resonances the correct relic abundance is obtained for couplings in
the perturbative range, $0.5\lesssim \lambda_\chi \lesssim 1$, and VEVs
between 200 and 500~GeV. Close to the resonances the enhancement of the
annihilation cross section has to be compensated by a large VEV (which
corresponds to small $\lambda_\chi$) in order to maintain the correct relic
abundance. Note that if the symmetry of the model was gauged, there would be
a $Z'$ gauge boson with a mass set by $w$. Since $Z'$ searches require the
mass of such a new gauge boson to be above few TeV,
e.g.~\cite{Carena:2004xs, Erler:2010uy, Accomando:2010fz, Basso:2010pe}, gauged versions
of this model would be confined to the resonance regions, see
ref.~\cite{Okada:2010wd} for an example of such a model. It is an advantage
of the global symmetry considered here, that the breaking scale can be
lower, and therefore this model does not suffer from the need of tuning the
DM mass close to half of the mass of one of the Higgs mass states.

\subsection{DM direct detection}
\label{sec:DD}

\begin{figure}
  \begin{center}
    \includegraphics[width=0.47\textwidth]{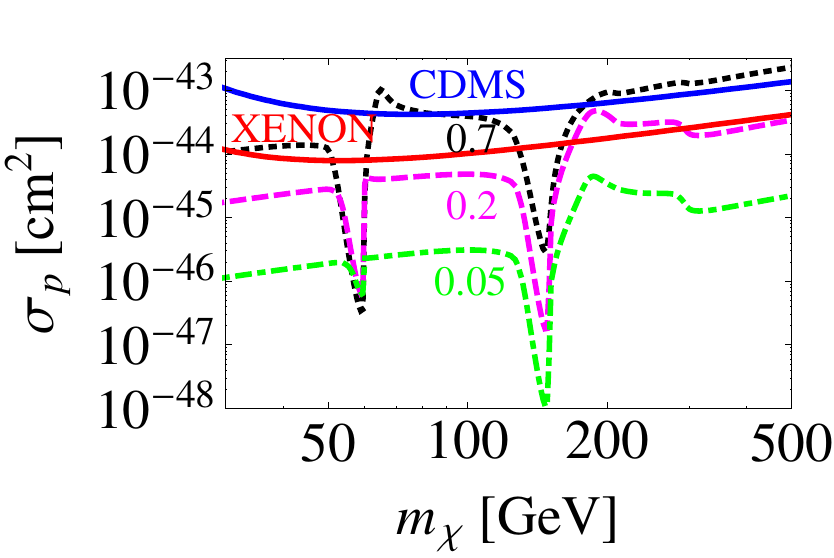}
    \includegraphics[width=0.47\textwidth]{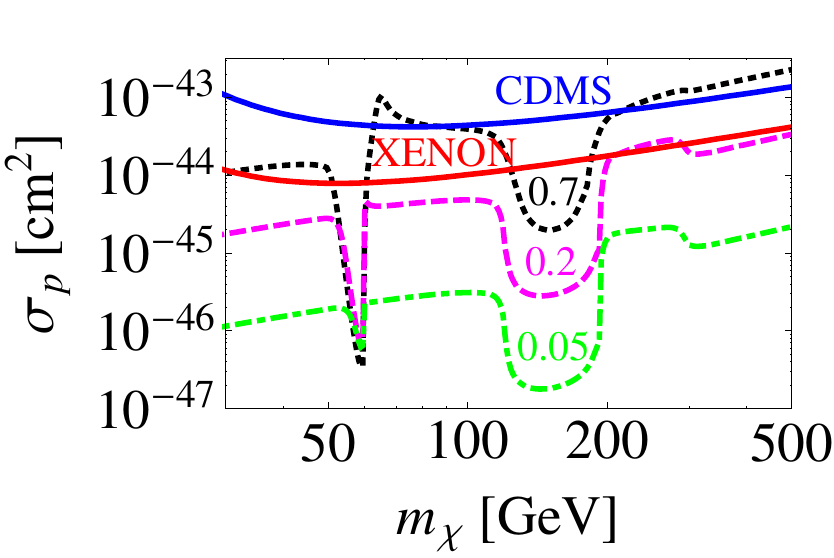} 
    \mycaption{Elastic scattering cross section $\sigma_p$ of $\chi$ off a
    proton $p$ for $m_{H_1}=300$ GeV and $m_{H_2}=120$ GeV and mixing angles
    $\sin\alpha=0.7,0.2,0.05$, according to the label.  The singlet VEV $w$
    has been chosen in order to obtain the correct relic DM abundance. The
    left (right) panel corresponds to masses of the Zee--Babu scalars of
    $k^{++} = 800 \,(400)$~GeV and $h^+ = 300\,(120)$~GeV. Also shown are
    the CDMS~\cite{Ahmed:2009zw} and XENON100~\cite{Aprile:2011hi} exclusion
    limits. } \label{fig:DD}
  \end{center}
\end{figure}

DM scattering on nuclei relevant for direct detection is mediated via
$t$-channel exchange of the Higgs mass eigenstates $H_1$ and $H_2$. Hence,
scattering is spin-independent. The elastic scattering cross section
$\sigma_{p}$ of $\chi$ off a proton $p$ is obtained as
\begin{equation}\label{eq:DD}
 \sigma_{p}=\frac{\lambda_\chi^2 \sin^22\alpha}{4 \pi} \, m_{\rm red}^{2} 
 \left(\frac{1}{m_{H_1}^2} - \frac{1}{m_{H_2}^2} \right)^{2} g_{Hp}^{2} \,,
\end{equation} 
where $m_{\rm red} = m_p m_\chi / (m_p + m_\chi)$ is the reduced mass of the DM--proton system and
\begin{equation}
 g_{Hp}=\frac{m_{p}}{v}\left[\sum_{q=u,d,s}f^{(p)}_{q}+\frac{2}{27}\left(1-
\sum_{q=u,d,s}f^{(p)}_{q}\right)\right]\,,
\end{equation}
see e.g., \cite{Belanger:2008sj} where also values for $f^{(p)}_q$ are
given. We observe from eq.~\eqref{eq:DD} that the cross section is
proportional to the Higgs mixing, since for $\alpha = 0$ the DM particle is
decoupled from the SM and therefore the scattering cross section will
vanish. Fig.~\ref{fig:DD} shows the DM cross section on a proton obtained
with micrOMEGAs~\cite{Belanger:2008sj}, where for given $m_{H_{1,2}},
\alpha$ and $m_\chi$ the coupling $\lambda_\chi$ (or equivalently the VEV
$w$) has been chosen such that the correct relic abundance is obtained.
Therefore the resonances in the annihilation cross section appear also in
the scattering cross section. Although the Zee--Babu scalars $k^{++}$ and
$h^+$ do not contribute directly to the scattering cross section, they
affect the annihilation cross section relevant for the relic abundance, and
therefore also DM--nucleus scattering depends indirectly on their masses, compare
left and right panel of fig.~\ref{fig:DD}. While part of the parameter space for large
values of $\sin\alpha$ is already excluded by present
bounds~\cite{Ahmed:2009zw, Aprile:2010um, Aprile:2011hi}, the cross section
can always be suppressed by making $\alpha$ small. Since generically mixing
should be sizable, one may expect observable signals in direct detection
searches from this model, although they are not guaranteed.

\subsection{DM indirect detection}
\label{sec:indir}

Let us comment briefly on possible signatures of our model for indirect searches for DM. The dominating branching fractions of the various DM annihilation channels are illustrated in fig.~\ref{fig:relic}. Generically our DM particles annihilate into a combination of gauge bosons and Higgs, leading to a rather typical mix of final annihilation products, with generic features common to many DM candidates produced thermally. In cases where DM annihilations into the Majoron dominate, visible annihilation products can be highly suppressed, which will lead to very small signals for indirect detection. 

If the charged scalars are not too heavy there might be a region in the parameter space where DM annihilates dominantly into $k^{\pm\pm}$ and/or $h^\pm$, which further will decay into charged leptons and neutrinos. In this case one might think that our model could explain anomalies in the cosmic electron/positron fluxes \cite{Adriani:2008zr,Ackermann:2010ij}, while respecting tight constraints for anti-protons. Note however, that a DM explanation of those excesses requires annihilation cross sections several orders of magnitude larger than the one needed for the thermal relic density. In our model no mechanism exists to enhance the cross section. Sommerfeld enhancement, for example, would require typically light (GeV-like) mediator particles, which are not present in our model. Therefore, attempting to explain those anomalies within our framework would require an extension of the model. However, those anomalies may very well be of astrophysical origin.


\subsection{DM self-interactions}
\label{sec:self}

As has been noted already in \cite{Chikashige:1980ui} the exchange of the
massless Majoron can lead to a long-range force. Hence, it is important to
check whether this leads to conflicts with bounds on DM self-interactions,
see e.g.~\cite{Ackerman:2008gi, Feng:2009mn, Buckley:2009in}. Due to the
pseudo-scalar nature of the Majoron, the interaction is spin-dependent and
the induced potential $V(r)$ falls off like $1/r^3$, in analogy to the
one--pion exchange potential in nuclear physics~\cite{Chikashige:1980ui,
Bedaque:2009ri}: 
\be
V(r) = \frac{3 \lambda_\chi^2}{16 \pi w^2 r^3} 
[3(\hat r \vec{\sigma_1})(\hat r \vec{\sigma_2}) -
\vec{\sigma_1}\vec{\sigma_2} ] \,,
\ee
where $\vec{\sigma}_{1}, \vec{\sigma}_{2}$ denote the spins of the two interacting particles.
Using analogous estimations as in~\cite{Ackerman:2008gi}, we have checked
that a potential of this type leads to less than one hard-scattering process
during the age of the Universe for a typical DM particle in the halo of a
Milky Way-like galaxy. Therefore, the Majoron induced self-interactions are
in agreement with the corresponding bounds.

\section{Discussion and conclusions}
\label{sec:discussion}

We have discussed in this Letter a common framework for neutrino mass and
Dark Matter where both are related to the breaking of a global $U(1)$
symmetry associated with the conservation of $B-L$.  From the perspective of
neutrino masses, we studied a modification of the Zee--Babu model, where
neutrinos obtain their masses at two-loop level after $U(1)_{B-L}$ breaking.
The single and double charged scalars of this model provide interesting
signatures at collider experiments as well as in searches for charged
lepton flavour violation~\cite{Babu:2002uu, AristizabalSierra:2006gb,
Nebot:2007bc}.  It is then tempting to use the same global $U(1)_{B-L}$
symmetry in order to stabilize Dark Matter. 
To provide a Dark Matter candidate we introduced two Majorana fermions with
non-standard $B-L$ quantum numbers, such that a Yukawa term with the SM
Higgs doublet is forbidden. They form a Dirac particle and are stable
because of an accidental $Z_2$ symmetry. Thanks to the non-standard quantum
numbers mentioned above in this framework the $Z_2$ symmetry emerges as an
unbroken remnant of $U(1)_{B-L}$ and has not to be imposed by hand. 

Due to the spontaneous breaking of the global $U(1)$ symmetry we obtain a
massless Goldstone boson, the Majoron. Quantum gravity effects are expected
to break global symmetries and hence Planck suppressed operators may provide
a mass to the Majoron. As discussed in~\cite{Akhmedov:1992hi} there are
certain constraints on operators from the requirement that the Majoron does
not over-close the 
Universe. On the other hand, if Majoron masses are induced
in the keV range, the Majoron could even provide a warm component to the
Dark Matter~\cite{Berezinsky:1993fm, Gu:2010ys} with interesting signatures due to
its decay~\cite{Bazzocchi:2008fh, Esteves:2010sh}. Here we have neglected
this contribution to Dark Matter, 
and furthermore, we have assumed that possible
Planck scale suppressed operators leading to DM decay maintain a lifetime of
the Dark Matter particle large compared to the age of the Universe.

There exists some literature on models for Dark Matter in the framework of
gauged $U(1)_{B-L}$, see e.g.~\cite{Okada:2010wd, Li:2010rb,
Nakayama:2011dj, Kanemura:2011vm, HernandezPinto:2011eu} and it is
interesting to compare with the scenario with a global $U(1)_{B-L}$
discussed here. In the gauged version the requirement of anomaly
cancellation provides a natural motivation to introduce right-handed
neutrinos while we had to postulate new fermions $N_i$.  In general the
right-handed neutrinos will have the usual Yukawa interactions with the SM
Higgs doublet, and hence, an additional $Z_2$ symmetry has to be postulated
if one (or more) of them should be the Dark Matter, while the protective
symmetry arises more naturally in our case.  Furthermore, a $Z'$ gauge boson
appears in the gauged version, which leads to additional constraints due to
$Z'$ searches at LEP, Tevatron, and LHC, e.g.~\cite{Carena:2004xs,
Erler:2010uy, Accomando:2010fz, Basso:2010pe}. For instance, from LEP-II
ref.~\cite{Carena:2004xs} finds $w = M_{Z'}/ (2 g_{B-L}) \gtrsim 3$~TeV.
Comparing this limit with the value of $w$ needed to obtain the correct Dark
Matter relic abundance in fig.~\ref{fig:relic}, we find that the collider
bound can be satisfied only close to the $s$-channel resonances, in
agreement with \cite{Okada:2010wd}. We have checked that also additional
contributions to the annihilation cross section due to $Z'$ exchange as well
as $Z'-B$ kinetic mixing cannot change this result.  Since the Dark Matter
mass and the scalar masses are unrelated in these models the requirement for
the resonance, $m_{H} \approx 2 m_\chi$, appears as an unnatural
coincidence. A recent example for a model with neutrino masses at one loop
and a gauged $U(1)$ broken to a remnant $Z_2$ has been given in
\cite{Chang:2011kv}.

Before concluding let us stress that the model discussed here may serve as a
specific example for a more general class of models, where Dark Matter is
charged under some global $U(1)$ symmetry and coupled to the Standard Model
via Higgs-mixing of a ``dark scalar'' whose VEV breaks the symmetry. Many
aspects of the phenomenology discussed here will apply to such models.
Furthermore, for models based on $B-L$, the neutrino mass mechanism is not
specific to the Zee--Babu model and one may think of various other
possibilities to generate neutrino masses due to the $U(1)_{B-L}$ breaking
linked to Dark Matter in the way described here. 


\vspace{5mm} \textbf{Acknowledgement.} We would like to thank Pei-Hong Gu and
Kristian McDonald for discussion. This work is supported by the
Sonderforschungsbereich TR 27 of the Deutsche Forschungsgemeinschaft
``Neutrinos and Beyond''. DS acknowledges support by the International Max
Planck Research School for Precision Tests of Fundamental Symmetries.

\bibliography{Babu_in_the_Dark_2}
\bibliographystyle{my-h-physrev.bst}

\end{document}